\documentclass[10pt,letterpaper]{article}
\usepackage[margin=1in]{geometry}
\usepackage{ifpdf}
\ifpdf
    \usepackage[pdftex]{graphicx}
    \graphicspath{{figs/}{matlab/}}   
    \usepackage[update]{epstopdf}
\else
	\usepackage{graphicx}
	\graphicspath{{figs/}{matlab/}}   
\fi
\usepackage{pdflscape}
\usepackage{marvosym}
\usepackage{enumitem,color}
\usepackage{amsthm}

\usepackage{hyperref}
\usepackage{array}
\usepackage{mathrsfs,bm,bbm}
\usepackage{amsmath,amsthm}
\usepackage{amsfonts}
\usepackage{amssymb}
\usepackage{cite,setspace}
\usepackage{breqn}
\usepackage{multirow}

\newtheorem{definition}{\textbf{Definition}}

\renewcommand{\vec}[1]{\mbox{\boldmath$#1$}}

\usepackage{xparse}
\NewDocumentCommand{\ml}{m O{M} O{Q_n^{[M]}}} {\mathcal{L}_{#1}(#2\rightarrow #3)}

\title{Broadcast Channel Cooperative Gain: An Operational Interpretation of Partial Information Decomposition}
\author{Chao Tian and Shlomo Shamai (Shitz)}

\begin{document}

\maketitle


\begin{abstract}
Partial information decomposition has recently found applications in biological signal processing and machine learning. Despite its impacts, the decomposition was introduced through an informal and heuristic route, and its exact operational meaning is unclear. In this work, we fill this gap by connecting partial information decomposition to the capacity of the broadcast channel, which has been well-studied in the information theory literature. We show that the synergistic information in the decomposition can be rigorously interpreted as the cooperative gain, or a lower bound of this gain, on the corresponding broadcast channel. This interpretation can help practitioners to better explain and expand the applications of the partial information decomposition technique.
\end{abstract}

\section{Introduction}

Shannon's mutual information has been widely accepted as a measure to gauge the amount of information that can be revealed by one random variable regarding another random variable. Partial information decomposition (PID) is an approach to refine and further decompose this fundamental quantity to explain the effect of interactions among several random variables. Recently, this approach has found applications in biological information processing \cite{wibral2017partial,feldman2024information,wibral2017quantifying,ehrlich2023measure,timme2014synergy,stramaglia2014synergy,timme2016high} and machine learning \cite{liang2024quantifying,liang2024foundations,liang2024factorized}. 
 
There exist different approaches to decompose the total information \cite{williams2010nonnegative,schneidman2003synergy,bertschinger2014quantifying,bertschinger2013shared,griffith2014quantifying,harder2013bivariate,olbrich2015information,lizier2018information,quax2017quantifying,barrett2015exploration,chatterjee2016construction,rauh2017extractable,ince2017measuring,finn2018pointwise,varley2024generalized}, but the general idea in the case with two observables is as follows: the total information revealed by $X$ and $Y$ regarding a third quantity $T$ is the mutual information between $(X,Y)$ and $T$, i.e., $I(X,Y;T)$, and it needs to be decomposed into four non-negative parts: 
\begin{itemize}
\item The common (or redundancy) information in $X$ and $Y$, regarding $T$;
\item The unique information in $X$, but not in $Y$, regarding $T$;
\item The unique information in $Y$, but not in $X$, regarding $T$;
\item The synergistic information (or complementary) information in $X$ and $Y$, regarding $T$, which only becomes useful when combined, but useless otherwise. 
\end{itemize}
This decomposition helps to explain the effect of combining $X$ and $Y$, or separately using $X$ or $Y$, to infer $T$. For example, in multi-modality machine learning, $X$ can represent one modality such as the vision image of the event, $Y$ another modality such as the soundtrack of the event, and $T$ is the event's label. In this case, we can use the amounts of synergistic and unique information to determine which multi-modality models will be the most effective in capturing the interactions and therefore most likely to be accurate. In neural signal processing, the quantities can be used in a similar manner to interpret different biological signals. 

One of the most influential notions of partial information decomposition was proposed by Bertschinger et al. \cite{bertschinger2014quantifying}, sometimes referred to as the BROJA PID (the initials of the authors)\footnote{We will simply refer to it as partial information decomposition as it is the only decomposition we consider.}. The desirable properties of this definition were thoroughly studied in \cite{bertschinger2014quantifying}, but in terms of the operation meaning,  only a qualitative justification in a decision-making setting was given. The intuition and motivation were that these quantities can help explain how rewards can be optimized via decision-making in such a setting, either using only $X$ (or only $Y$) as the observations, or using $X$ and $Y$ jointly as the observations. However, this qualitative justification is far from satisfying, as it does not provide the exact quantitative interpretation, and the meaning is quite vague. In this work, we fill this gap by connecting PID to the well-studied broadcast channel in information theory \cite{cover1972broadcast,cover1975achievable,weingarten2006capacity}, and show that there is indeed a quantitative connection between PID and the sum-rate capacity of the broadcast channel. More precisely, the synergistic information defined in \cite{bertschinger2014quantifying} can be interpreted as either the cooperative gain or a lower bound of the cooperative gain in the broadcast channel, under a fixed signaling distribution.

\section{Partial Information Decomposition}

\subsection{A Few Examples}
Before formally introducing the PID defined in \cite{bertschinger2014quantifying}, let us consider a few simple settings to understand intuitively how the decomposed information should behave. 
\begin{enumerate}
\item Common information dominant: $X$ is a uniformly distributed Bernoulli random variable and $Y=T=X$. It is clear that $I(X,Y;T)=1$. In this case, the common information should be $1$, since $X$ and $Y$ are exactly identical. The other components should all be $0$, since there is no unique information in either $X$ or $Y$ regarding $T$, and there is no synergistic information when combining $X$ and $Y$. 
\item Synergistic information dominant: $X$ and $Y$ are uniformly distributed Bernoulli random variables independent of each other, and $T=X \oplus Y$, where $\oplus$ is the XOR operation. Here, the synergistic information should be $1$, and the other components should be $0$. This is because $X$ or $Y$ alone do not reveal any information regarding $T$, and since they are completely independent, they do not share any common information either, but their combination reveals the full information on $T$. 
\item Component-wise decomposition: $X=(X_1,X_2)$ and $Y=(Y_1,Y_2)$, where $X_1,X_2,Y_1,Y_2$ are all uniformly-distributed Bernoulli random variables, mutually independent of each other. Let $T=(X_1,Y_1,X_2 \oplus Y_2)$. Clearly, in this case, the common information is still $0$, but the two kinds of unique information are both $1$, and the synergistic information is also $1$.
\end{enumerate}

\subsection{Partial Information Decomposition}
We next introduce the formal definition for the partial information decomposition given in \cite{bertschinger2014quantifying}. Let $X,Y,T$ be three random variables in their respective alphabet $\mathcal{X}$, $\mathcal{Y}$, and $\mathcal{T}$, which follow the joint distribution $P_{X,Y,T}$. The total information between $X,Y$ and $T$ is written as $I_P(T;X,Y)$, where the subscript $P$ is used to emphasize the distribution that the mutual information is computed for. As mentioned earlier, the decomposition first needs to satisfy the total information rule: 
\begin{align}
&\text{\textbf{Total information:}\,\,\,\,\,}\notag\\
&\qquad I_P(T;X,Y)= I^{(C)}(T;X,Y)+I^{(U_X)}(T;X,Y)+I^{(U_Y)}(T;X,Y)+I^{(S)}(T;X,Y),
\label{eqn:total}
\end{align}
where the superscript $C$ indicates common (or redundancy) information, $U_X$ indicates unique information from $X$, $U_Y$ indicates unique information from $Y$, and $S$ indicates synergistic (or complementary) information. This condition simply states that the total information is the summation of four parts.
The decomposition also needs to satisfy the individual information rule: 
\begin{align}
\text{\textbf{Individual information:}\,\,\,\,\,}
\begin{array}{ll}
I_P(T;X)=I^{(C)}(T;X,Y)+I^{(U_X)}(T;X,Y)\\
I_P(T;Y)=I^{(C)}(T;X,Y)+I^{(U_Y)}(T;X,Y).
\label{eqn:individual}
\end{array}
\end{align}
This rule dictates that the information that can be revealed from $X$ regarding $T$ needs to be the summation of the unique information in $X$, and the common information in both $X$ and $Y$. Note that the synergistic information is not included here, since it can only manifest when combining $X$ and $Y$.

We now already have three linear equations for the four quantities in the decomposition, and the authors of \cite{bertschinger2014quantifying} introduced the following definition of synergistic information:
\begin{align}
I^{(S)}(T;X,Y)\triangleq I_P(T;X,Y)-\min_{Q\in\mathcal{Q}} I_Q(T;X,Y),\label{eqn:syn}
\end{align}
where the set of distributions $\mathcal{Q}$ is defined as 
\begin{align}
\mathcal{Q}=\{Q\in \Delta_{\mathcal{X}\times\mathcal{Y}\times\mathcal{T}}:Q_{{T},{X}}=P_{T,X}, Q_{{T},{Y}}=P_{T,Y}\},
\label{eqn:setQ}
\end{align}
where $\Delta_{\mathcal{X}\times\mathcal{Y}\times\mathcal{T}}$ is the probability simplex on $\mathcal{X}\times\mathcal{Y}\times\mathcal{T}$; in other words, $\mathcal{Q}$ is the set of all joint distributions, such that the pairwise marginals of $P_{T,X}$ and $P_{T,Y}$ are preserved. 
As a consequence, this definition, together with (\ref{eqn:total}) and (\ref{eqn:individual}), also specifies the values of the other three quantities $I^{(C)}(T;X,Y)$, $I^{(U_X)}(T;X,Y)$, and $I^{(U_Y)}(T;X,Y)$. It can be easily verified that in the examples discussed earlier, this decomposition gives the quantities matching our expectations. 

It was shown in \cite{bertschinger2014quantifying} that this PID definition enjoys several desirable properties, for example: 1) All the quantities $I^{(C)}(T;X,Y)$, $I^{(U_X)}(T;X,Y)$, $I^{(U_Y)}(T;X,Y)$, and $I^{(S)}(T;X,Y)$ are non-negative; 2) They satisfy certain extremal relations in the sense that the shared information $I^{(S)}(T;X,Y)$ is the minimum among all possible definition under an additional axiom that the unique information only depends on the two marginal distributions; 3) They satisfy the symmetry, self-redundancy, and monotonicity axioms stipulated in \cite{williams2010nonnegative}. 

\section{Broadcast Channels}

\begin{figure}[t!]
\includegraphics[width=1.0\textwidth]{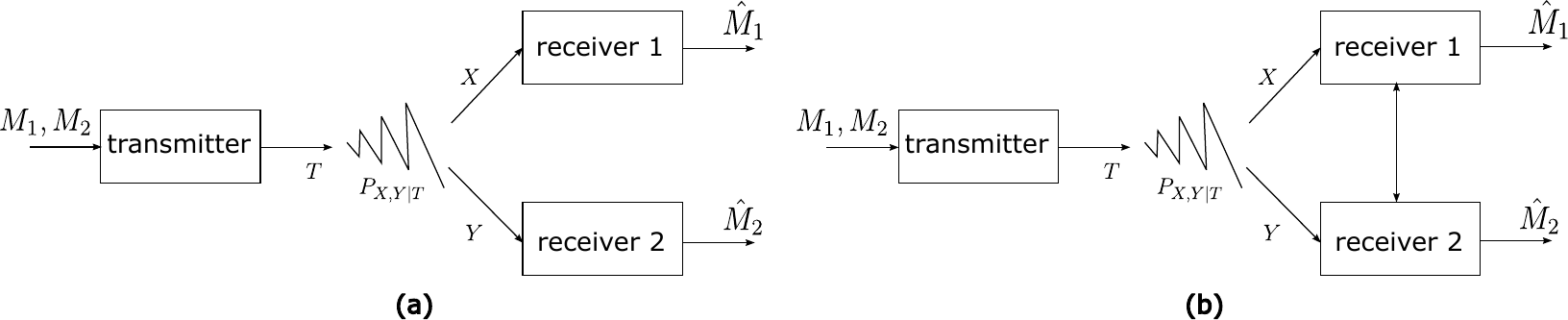}
\caption{Standard broadcast channel vs. broadcast channel with receiver cooperation. \label{fig:BC} }
\end{figure}

The broadcast channel is a well-studied communication system in classic information theory \cite{cover1972broadcast}, where a transmitter wishes to send two independent messages to two receivers through a channel with a single input signal that induces two separate output signals; see Fig. \ref{fig:BC} (a). To accomplish this goal, the transmitter must encode the messages in such a way that the individual messages can be decoded by the intended receivers. Clearly, if the first receiver is completely ignored, the transmitter can send more information to the second receiver, and vice versa. In other words, there is a tradeoff between the amounts of information that can be sent to the two receivers. One particular important quantity here is the sum of the rates that can be supported on this channel, often referred to as the sum-rate capacity.

Next, we provide a rigorous definition of the broadcast channel capacity region, where the channel is allowed to be used many times at a time, i.e., a block code where the block length approaches infinity. A \emph{two-user broadcast channel} is specified by an input alphabet $\mathcal{T}$, two output alphabets $\mathcal{X}$ and $\mathcal{Y}$, and a conditional probability distribution $P_{X,Y\mid T}$ that gives the channel law for each symbol $t \in \mathcal{T}$ and $(x,y) \in \mathcal{X} \times \mathcal{Y}$. The alphabets are usually assumed to be finite, though the results usually hold under more general assumptions \cite{el2011network}. 

\begin{definition}    
For a blocklength $n \in \mathbb{N}$, let $M_1$ and $M_2$ be two positive 
integers. An $(n, M_1, M_2)$-code for the broadcast channel 
$P_{X,Y\mid T}$ consists of:
\begin{itemize}
    \item Two message sets: 
    \[
      \mathcal{M}_1 = \{1,2,\ldots,M_1\}, 
      \quad 
      \mathcal{M}_2 = \{1,2,\ldots,M_2\},
    \]
    \item An encoding function 
    \[
       f : \mathcal{M}_1 \times \mathcal{M}_2 \to \mathcal{T}^n,
    \]
    which assigns to each pair of messages $(m_1,m_2)$ a length-$n$ input 
    sequence $T^n = (T_1,T_2,\ldots,T_n)$,
    \item Two decoding functions,
    \[
      g_1 : \mathcal{X}^n \to \mathcal{M}_1, 
      \quad 
      g_2 : \mathcal{Y}^n \to \mathcal{M}_2,
    \]
    where $\hat{M}_1=g_1(X^n)$ is the estimate of message $M_1$ by receiver 1, and 
    $\hat{M}_2=g_2(Y^n)$ is the estimate of message $M_2$ by receiver 2.
\end{itemize}
\end{definition}

\begin{definition}
For an $(n,M_1,M_2)$-code, let $(M_1,M_2)$ be chosen uniformly from 
$\mathcal{M}_1 \times \mathcal{M}_2$. The average probability of error is defined as
\[
  P_e^{(n)} 
  = \Pr\bigl\{(\hat{M}_1,\hat{M}_2) \neq (M_1,M_2)\bigr\}.
\]
A sequence of codes (indexed by $n$) is said to have vanishing error probability if $P_e^{(n)} \to 0$ as $n \to \infty$.
\end{definition}

\begin{definition}
A rate pair $(R_1, R_2)$ is said to be \emph{achievable} for the broadcast 
channel $P_{X,Y\mid T}$ if there exists a sequence of 
$\bigl(n, M_1^{(n)}, M_2^{(n)}\bigr)$-codes with vanishing error probability 
such that 
\[
  \liminf_{n \to \infty} \frac{1}{n}\log M_1^{(n)} \ge R_1,
  \quad
  \liminf_{n \to \infty} \frac{1}{n}\log M_2^{(n)} \ge R_2.
\]
The capacity region $\mathcal{C}$ of the two-user broadcast channel 
is the closure of the set of all achievable rate pairs. The sum-rate capacity is $C_{\text{sum}}\triangleq\max_{(R_1,R_2)\in \mathcal{C}} R_1+R_2$.
\end{definition}

\section{Main Result: An Operational Interpretation of PID}

In this section, we provide the main result of this work, which is an operational interpretation of partial information decomposition. 

\subsection{PID via Sato's Outer Bound}

Researchers in the information theory community made numerous efforts to identify a computable characterization of the capacity region of general broadcast channels (see the textbook \cite{el2011network} for a historical summary), yet at this time, a complete solution is still elusive. Nevertheless, significant progress has indeed been made toward this goal. Particularly, Sato \cite{sato1978outer} provided an outer bound for $\mathcal{C}$, and it can be specialized to yield an upper bound for the sum-rate capacity of the general broadcast channel as follows:
\begin{align}
C_{\text{sum}}\leq C_{\text{Sato}}\triangleq\min_{Q\in\mathcal{Q'}}\max_{P_T} I_{P_TQ_{X,Y\mid T}}(X,Y; T),\label{eqn:sato}
\end{align}
where the set $\mathcal{Q}'$ is defined as follows
\begin{align}
\mathcal{Q}'=\{Q\in \Delta_{\mathcal{X}\times\mathcal{Y}\mid \mathcal{T}}:Q_{{X}|{T}}=P_{X|T}, Q_{{Y}|T}=P_{Y|T}\},
\label{eqn:setQprime}
\end{align}
i.e., the set of conditional distributions that the marginal conditional distributions $P_{X|T}$ and $P_{Y|T}$ are preserved. The inner maximization is over the possible marginal distribution of the random variable $P_T$ in the alphabet $\mathcal{T}$. The form already bears a certain similarity to (\ref{eqn:setQ}). Note that for channels on general alphabets (i.e., not necessarily optimized on a compact space), the maximization should be replaced by the supremum and the minimization by the infimum. Due to the minimax form, the meaning is not yet clear, but the max-min inequality implies that 
\begin{align}
\min_{Q\in\mathcal{Q'}}\max_{P_T} I_{P_TQ_{X,Y\mid T}}(X,Y; T)\geq \max_{P_T} \min_{Q\in\mathcal{Q'}} I_{P_TQ_{X,Y\mid T}}(X,Y; T)=\max_{P_T} \min_{Q\in\mathcal{Q}} I_{Q_{X,Y, T}}(X,Y; T),
\end{align}
where the set of $\mathcal{Q}$ is exactly the one defined in (\ref{eqn:setQ}) with $Q_{X,T}=P_TQ_{X|T}$ and $Q_{Y,T}=P_TQ_{Y|T}$. The inner minimization of this form is exactly the same as the second term in (\ref{eqn:syn}). Though the max-min form does not yield a true upper bound of the sum-rate capacity, in the PID setting we consider, $P_T$ is always fixed, therefore, the max-min and min-max forms are in fact equivalent in this setting. The equivalence in mathematical forms does not fully explain the significance of this connection, and we will need to consider Sato's bound more carefully. Let us define the following quantity for notational simplicity 
\begin{align}
R_{P_T}\triangleq \min_{Q\in\mathcal{Q}'} I_{P_TQ_{X,Y\mid T}}(X,Y; T).
\end{align}

Sato's outer bound was derived using the following argument. For a channel that has a single input signal $T$ and a single output signal $Y$ on a channel $P_{Y|T}$, Shannon's channel coding theorem \cite{cover1999elements} states that the channel capacity is given by $\max_{P_T} I_{P_TP_{Y|T}}(Y;T)$. Moreover, for any fixed distribution $P_T$, the rate $I_{P_TP_{Y|T}}(Y;T)$ is achievable, where the probability distribution $P_T$ represents the statistical signaling pattern of the underlying codes. Turning our attention back to the broadcast channel with a transition probability $P_{X,Y|T}$, if the receivers are allowed to cooperate fully and share the two output signals $X$ and $Y$, i.e., they become a single virtual receiver (see Fig. \ref{fig:BC} (b)), then clearly the maximum rate achievable would be $\max_{P_T} I_{P_TQ_{X,Y\mid T}}(X,Y;T)$. However, Sato further observed that the error probability of any code should not differ on any broadcast channel $Q_{X,Y\mid T}\in \mathcal{Q}$, even if the transition distribution $Q_{X,Y\mid T}$ is different from the true broadcast channel transition probability $P_{X,Y\mid T}$. This is because the channel outputs only depend on the marginal transition probabilities $P_{X|T}$ and $P_{Y|T}$, respectively, and the decoders only use their respective channel outputs to decode. Therefore, we can obtain an upper bound by choosing the worst channel configuration $Q\in \mathcal{Q}$, i.e., the outer minimization in (\ref{eqn:sato}).

With the interpretation of Sato's upper bound above, it becomes clear that $R_{P_T}$ is essentially an upper bound on the sum rate of the broadcast channel where the receivers are not allowed to cooperate, when the input signaling pattern is fixed to follow $P_T$. On the other hand, the quantity $I_{P_TP_{X,Y\mid T}}(X,Y;T)$ is the rate that can be achieved by allowing the two receivers to fully cooperate, also with $P_T$ being the input signaling pattern. In this sense, $I^{(S)}(T;X,Y)$ defined in (\ref{eqn:syn}) is a lower bound on the difference between the sum rate with full cooperation and that without any cooperation, with the input signaling pattern following $P_T$. 

This connection provides an operational interpretation of PID for general distributions. Essentially, synergistic information can be viewed as a surrogate of the cooperative gain. When this lower bound is in fact also achievable, $I^{(S)}(T;X,Y)$ would be exactly equal to the cooperative gain. In the corresponding learning setting, it is the difference between what can be inferred about $T$ by using both $X$ and $Y$ in a non-cooperative manner, and what can be inferred by using them jointly. This indeed matches our expectations for the synergistic information. In the next subsection, we consider a special case when Sato's bound is indeed achievable, and the lower bound mentioned above becomes exact.

In one sense, this operational interpretation is quite intuitive as explained above, but on the other hand, it is also quite surprising. For example, in broadcast channels, a more general setup allows the transmitter to also send a common message to both receivers \cite{geng2014capacity,tian2011latent}, in addition to the two individual messages to the two respective receivers. It would appear plausible to expect this generalized setting to be more closely connected to the PID setting with the common message related to the common information, yet this turns out to be not the case here. Moreover, a dual communication problem studied in information theory is the multiple access channel (see e.g., \cite{cover1999elements}), where two transmitters wish to communicate to the same receiver simultaneously. The readers may also wonder if an operational meaning should be extracted on this channel, instead of on the broadcast channel. However, note that in the PID setting, we are inferring $T$ from $X$ and $Y$, which is similar to the decoding process in the broadcast channel, instead of the multiple access channel. Moreover, in the multiple access channel, the joint distribution of the two transmitters' inputs is always independent when the two transmitters cannot cooperate, and this will not match the PID setting in consideration. Another seemingly related problem setting studied in the information theory literature is the common information between two random variables \cite{gacs1973common, wyner1975common}, however, for the PID defined in \cite{bertschinger2014quantifying}, this approach also does not yield a meaningful interpretation.

\subsection{Gaussian MIMO Broadcast Channel and Gaussian PID}

One setting where a full capacity region characterization is indeed known is the Gaussian multiple-input multiple-output (MIMO) channel \cite{yu2004sum,weingarten2006capacity}. In the two-user Gaussian MIMO broadcast channel, the channel transition probability $P_{\vec{X}|\vec{T}}$ and $P_{\vec{Y}|\vec{T}}$ are given, with $\vec{T}$ being the transmitter input variable, and $\vec{X},\vec{Y}$ the channel outputs at the two individual receivers. The channel is usually defined as follows: 
\begin{align}
&\vec{X}={H}_X \vec{T}+\vec{n}_X\\
&\vec{Y}={H}_Y \vec{T}+\vec{n}_Y,
\end{align}
where ${H}_X$ and ${H}_y$ are two channel matrices, the additive noise vector $\vec{n}_X$ is independent of $\vec{T}$, and similarly  $\vec{n}_Y$ is independent of $\vec{T}$. For a fixed input signaling distribution $P_T$, the pairwise marginal distributions $P_{\vec{T},\vec{X}}$ and $P_{\vec{T},\vec{Y}}$ are well-specified. Conversely, for any joint distribution $P_{\vec{T},\vec{X},\vec{Y}}$, where the marginals $P_{\vec{T},\vec{X}}$ and $P_{\vec{T},\vec{Y}}$ jointly Gaussian, respectively, we can represent their relation in the form above via a Gram–Schmidt orthogonalization. Note that the joint distribution of $\vec{n}_X,\vec{n}_Y$ is not fully specified here, as the noise vectors $\vec{n}_X$ and $\vec{n}_Y$ are not necessarily jointly Gaussian, but can be dependent in a more sophisticated manner. The standard Gaussian MIMO broadcast problem usually specifies the noises zero-mean Gaussian with certain fixed covariances, and there is also a covariance constraint on the transmitter's signaling $\vec{T}$. The problem can be further simplified using certain linear transformations, as discussed below. 

Let us assume $\Sigma_{\vec{n}_X}$ and $\Sigma_{\vec{n}_Y}$ are full rank for now\footnote{When $\Sigma_{\vec{n}_X}$ and $\Sigma_{\vec{n}_Y}$ are not fully rank, a limiting argument can be invoked to show the same conclusion holds. }, and in this case, it is clearly without loss of generality to assume that $\Sigma_{\vec{n}_i}$'s are in fact identity matrices, since otherwise, we can perform receiver side linear transforms to make them so, i.e., through a transformation based on the eigenvalue decomposition of $\Sigma_{\vec{n}_X}$ and $\Sigma_{\vec{n}_Y}$, respectively. For the same reason, we can assume $\Sigma_{\vec{T}}$ to an identity matrix, through a linear transformation at the transmitter. These reductions to independent noise and independent channel input are often referred to as the transmitter precoding transformation and the receiver precoding transformations in the communication literature; see e.g., \cite{yu2004sum}. 

For the two-user Gaussian broadcast channel, the worst channel configuration problem we discussed in the general setting is essentially the least favorable noise problem considered by Yu and Cioffi \cite{yu2004sum} with the simplification above, where the noise relation between $\vec{n}_X$ and $\vec{n}_Y$ needs to be identified for a channel that makes it the hardest to communicate. It was shown in \cite{yu2004sum} that the least favorable noise problem can be recast as an optimization problem: 
\begin{align}
\text{minimize: }& \log\frac{|H\Sigma_{\vec{T}}H^T+\Sigma_{\vec{n}}|}{|\Sigma_{\vec{n}}|}\label{eqn:obj}\\
\text{subject to: }& \Sigma_{\vec{n}_x}=I,\\
                  & \Sigma_{\vec{n}_y}=I,\\
                 &\Sigma_{\vec{n}}\succeq 0,
                 \label{eqn:const}
\end{align}
where $H^T=[H_X^T,H_Y^T]$, and $\vec{n}=(\vec{n}_X,\vec{n}_Y)$, when $H\Sigma_{\vec{T}}H^T$ is nonsingular. It can be shown that the problem is convex. 

Yu and Cioffi also showed that in this setting, Sato's upper bound is achievable, i.e., it is exactly the sum-rate capacity. Moreover, for any input signaling $P_{\vec{T}}$ that is Gaussian distributed, the corresponding $R_{P_T}$ can be achieved on this broadcast channel through a more sophisticated scheme known as dirty-paper coding \cite{costa1983writing}. Therefore, when the pairwise marginals $P_{\vec{T},\vec{X}}$ and $P_{\vec{T},\vec{X}}$ are jointly Gaussian, respectively, the synergistic information $I^{(S)}(\vec{T};\vec{X},\vec{Y})$ is exactly the cooperative gain of the corresponding Gaussian broadcast channel using this specific input signaling pattern $P_{\vec{T}}$. The connection in the Gaussian setting is of particular interest, given the practical importance of the Gaussian PID, which was thoroughly explored in \cite{venkatesh2024gaussian}.

\subsection{Revisiting the Examples}

Let us now revisit the examples given earlier, and attempt to understand them in the broadcast channel setting:
\begin{enumerate}
\item Common information dominant: Here the two receivers both know the transmitted signal completely, and therefore, there is no difference even if they are allowed to cooperate, and the cooperative gain is 0. 
\item Synergistic information dominant: Clearly here $T$ is a uniformly distributed Bernoulli random variable. When the two receivers are not allowed to cooperate, they cannot decode anything because the channel $T\rightarrow X$ is completely noisy, and similarly for the channel $T\rightarrow Y$. When the receivers are allowed to cooperate, then from $X$ and $Y$, we can completely recover $T$, i.e., the channel becomes noiseless, therefore, the full information in $T$ can be decoded. The cooperative gain is then 1, matching the synergistic information. 
\item Component-wise decomposition: Here $T$ has three uniformly distributed Bernoulli random variable components, mutually independent of each other. In the channel $T\rightarrow X$, the first component is noiseless, and the second component is completely noisy. Similarly, for $T\rightarrow Y$. When the receivers are not allowed to cooperate, the sum rate on the channel is $2$. However, when the channels are allowed to cooperate, the last component in $T$ becomes useful (noiseless), and the total communication rate becomes $3$. Therefore the cooperative gain is $3-2=1$, equal to the synergistic information.
\end{enumerate}

\section{Conclusion}
We provide an operational interpretation of the partial information decomposition given in \cite{bertschinger2014quantifying} via a connection to the well-studied broadcast channel capacity in the information theory literature. The synergistic information is directly connected to the cooperative gain on the corresponding broadcast channel, either quantitatively exactly equal or being a lower bound. This interpretation can help us understand better why such decomposition can be used to guide model selection in machine learning and in the analysis of biological signals, and potentially design new methods to utilize it. 

We note that in the information theory literature, broadcast channels with cooperative receivers have been studied carefully \cite{dabora2006broadcast}, where the two receivers are allowed to conference on rate-limited links. It may be of interest to consider the corresponding notion of partial information decompositions, when the synergistic information can be parametrized by the degree of cooperation between the receivers. 

\bibliographystyle{IEEEtran}
\bibliography{PID}
\end{document}